\documentclass{ifacconf}

\usepackage{graphicx}      
\usepackage{natbib}        

\usepackage[utf8]{inputenc}
\usepackage[T1]{fontenc}
\usepackage{graphicx}
\usepackage{amsmath}
\usepackage{amssymb}
\usepackage{amsfonts}
\usepackage{bm}
\usepackage{hhline}
\usepackage{csquotes}
\usepackage{color}
\usepackage{import}
\usepackage{algorithm}
\usepackage{algpseudocode}
\usepackage{dsfont}
\usepackage{multirow}
\usepackage{tensor}
\usepackage{soul}

\usepackage{url}

\newcommand{\mbf}[1]{\ensuremath{{\mathbf{#1}}}}

\newcommand{\eye}[1]{\ensuremath{\mbf{I}_{#1}}}
\newcommand{\eyenoarg}{\ensuremath{\mbf{I}}}
\newcommand{\zeros}[2]{\ensuremath{\bm{0}_{#1\times#2}}}

\newcommand{\real}[1]{\ensuremath{\text{Re}(#1)}}

\newcommand{\realset}{\ensuremath{\mathbb{R}}}
\newcommand{\realsetmat}[2]{\ensuremath{\mathbb{R}^{#1\times#2}}}

\newcommand{\ninf}[1]{\ensuremath{\|{#1}\|_\infty}}

\makeatletter
\newcommand{\oset}[2]{%
   {\mathop{#2}\limits^{\vbox to -.5\ex@{\kern-\tw@\ex@
    \hbox{\scriptsize #1}\vss}}}}
\makeatother

\newcommand{\seq}[1]{\ensuremath{{\overset{\to}{#1}}}}

\newcommand{\zerospace}{\ensuremath{{\,\!}}}
\newcommand{\inv}[1]{\ensuremath{{#1}\zerospace^{-1}}}

\newtheorem{theorem}{Theorem}
\newtheorem{lemma}{Lemma}
\newtheorem{definition}{Definition}
\newtheorem{remark}{Remark}
\newtheorem{assumption}{Assumption}

\allowdisplaybreaks 

\newcommand{\textred}[1]{#1}
\newcommand{\textgreen}[1]{#1}
\newcommand{\textblue}[1]{#1}

\begin{document}
\begin{frontmatter}


\title{Set-based state estimation and fault diagnosis of linear discrete-time descriptor systems using constrained zonotopes\thanksref{footnoteinfo}\vspace{-1mm}} 

\thanks[footnoteinfo]{This work has been supported by the Italian Ministry for Research in the framework of the 2017 PRIN, Grant no. 2017YKXYXJ,
the brazilian project INCT under the grant CNPq 465755/2014-3, FAPESP 2014/50851-0, and also by the brazilian agencies CAPES under the grants 001 and 88887.136349/2017-00, and FAPEMIG. \copyright 2020 the authors. This work has been accepted to IFAC for publication under a Creative Commons Licence CC-BY-NC-ND.} 

\author[First]{Brenner S. Rego} 
\author[Second]{Davide M. Raimondo} 
\author[First,Third]{Guilherme V. Raffo}


\address[First]{Graduate Program in Electrical Engineering, Federal University of Minas Gerais, Belo Horizonte, Brazil (e-mail: brennersr7@ufmg.br)}
\address[Second]{Department of Electrical, Computer and Biomedical Engineering, University of Pavia, Pavia, Italy (e-mail: davide.raimondo@unipv.it)}
\address[Third]{Department of Electronics Engineering, Federal University of Minas Gerais, Belo Horizonte, Brazil (e-mail: raffo@ufmg.br)}

\begin{abstract}
This paper presents new methods for set-valued state estimation and active fault diagnosis of linear descriptor systems. The algorithms are based on constrained zonotopes, a generalization of zonotopes capable of describing strongly asymmetric convex sets, while retaining the computational advantages of zonotopes. Additionally, unlike other set representations like intervals, zonotopes, ellipsoids, paralletopes, among others, linear static constraints on the state variables, typical of descriptor systems, can be directly incorporated in the mathematical description of constrained zonotopes. Therefore, the proposed methods lead to more accurate results in state estimation in comparison to existing methods based on the \textblue{previous} sets without requiring rank assumptions on the structure of the descriptor system and with a fair trade-off between accuracy and efficiency. These advantages are \textblue{highlighted} in two numerical examples. 
\end{abstract}

\end{frontmatter}

\section{Introduction}


Many physical processes such as battery packs, robotic systems with holonomic and nonholonomic constraints, and socioeconomic systems \citep{JAN11,Yang2019}, exhibit static relations between their internal variables. These processes are known as descriptor systems (or implicit systems), \textblue{which have generalized dynamic and static behaviors described through differential and algebraic equations, respectively~\citep{Puig2018}}. Descriptor systems appear in several contexts, such as linear control \citep{Laub1987}, fault-tolerant control \citep{Shi2014}, and fault diagnosis \citep{Wang2019b}. \textblue{However, few strategies can deal effectively with state estimation and fault diagnosis of descriptor systems when uncertainties with unknown probability distribution are present} \citep{Hamdi2012}.

Set-based strategies for state estimation of discrete-time descriptor systems with unknown-but-bounded uncertainties is a recent subject, often addressed using intervals, zonotopes and ellipsoids \citep{Efimov2015,Puig2018,Wang2018,Merhy2019}. Interval methods are used in \cite{Efimov2015} to design a state estimator for time-delay descriptor systems based on linear matrix inequalities and the Luenberger structure. However, interval arithmetics can lead to conservative enclosures due to the wrapping effect. A few methods are proposed in \cite{Puig2018} for set-valued state estimation of linear descriptor systems by enclosing the intersection of two consistent sets with a zonotope bound. Nevertheless, since the intersection cannot be computed exactly, the resulting bound can be conservative. These strategies are extended in \cite{Wang2018} for linear parameter-varying descriptor systems, but conservative enclosures are still present since the intersection method is maintained. Ellipsoids are used in \cite{Merhy2019} for state estimation of linear descriptor systems based on Luenberger type observers. Despite being able to provide stable bounds, the ellipsoidal estimation can be conservative since the complexity of the set is fixed and static relations are not directly incorporated. In addition, as in \cite{Puig2018}, restrictive assumptions on the rank of the system matrices are required to be able to design the proposed estimator. Moreover, due to the static constraints, the \textblue{reachable} sets of models of a descriptor system may be asymmetric even if the initial set is symmetric, and therefore methods based on the sets above can provide conservative enclosures.

Fault diagnosis aims to determine exactly which fault a process is subject to. For the case of descriptor systems, this problem has been considered in recent works using zonotopes \citep{Yang2019,Wang2019b}. While \cite{Wang2019b} explores the use of unknown input observers for robust passive fault diagnosis limited to additive faults, \cite{Yang2019} proposes a zonotope-based method for active fault diagnosis (AFD) of descriptor systems. The latter is based on the design of an input sequence for separation of the reachable sets. Unfortunately, the generator representation of zonotopes cannot incorporate exactly static relations between the state variables in general linear descriptor systems. Consequently, in practice, this may lead to a more difficult diagnosis.

This manuscript proposes strategies for set-based state estimation and AFD of linear descriptor systems based on constrained zonotopes (CZs), aiming to reduce the conservativeness described above through this asymmetric set representation. In contrast to the aforementioned sets, it is worth noting that linear static constraints, typical of descriptor systems, can be directly incorporated in the mathematical description of CZs. Thanks to this feature, set-valued methods based on CZs can provide less conservative enclosures. The proposed methods extensively explore the basic properties of CZs \citep{Scott2016}. In addition, efficient complexity reduction methods for CZs, available in the literature \citep{Scott2016,Scott2018}, are used in the proposed strategies allowing a direct trade-off between accuracy and computational efficiency. The superiority of the proposed approaches with respect to zonotope-based methods is highlighted in numerical examples. 



\subsection{Problem formulation and preliminaries}

Consider a linear discrete-time descriptor system with time $k$, state $\mbf{x}_k \in \realset^{n}$, input $\mbf{u}_{k} \in \realset^{n_u}$, process uncertainty $\mbf{w}_k \in \realset^{n_w}$, measured output $\mbf{y}_k \in \realset^{n_y}$, and measurement uncertainty $\mbf{v}_k \in \realset^{n_v}$. In each time interval $[k-1,k]$, $k=1,2,\ldots$, the system evolves according to the model
\normalsize
\begin{equation}
\begin{aligned} \label{eq:system}
\mbf{E} \mbf{x}_k & = \mbf{A} \mbf{x}_{k-1} + \mbf{B} \mbf{u}_{k-1} + \mbf{B}_w \mbf{w}_{k-1}, \\
\mbf{y}_k & = \mbf{C} \mbf{x}_k + \mbf{D} \mbf{u}_{k} + \mbf{D}_v \mbf{v}_{k},
\end{aligned}
\end{equation}
\normalsize
with $\mbf{E} \in \realsetmat{n}{n}$, $\mbf{A} \in \realsetmat{n}{n}$, $\mbf{B} \in \realsetmat{n}{n_u}$, $\mbf{B}_w \in \realsetmat{n}{n_ w}$, $\mbf{C} \in \realsetmat{n_y}{n}$, $\mbf{D} \in \realsetmat{n_y}{n_u}$, and $\mbf{D}_v \in \realsetmat{n_y}{n_v}$. In the following, it is assumed that $\text{rank}(\mbf{E}) \leq n$. Note that when $\mbf{E}$ is singular, one has $n - \text{rank}(\mbf{E})$ purely static constraints. It is assumed that the initial state $\mbf{x}_0 \in X_0$ and $(\mbf{w}_{k},\mbf{v}_{k}) \in W \times V$ for all $k \geq 0$, where $X_0$, $W$ and $V$ are known polytopic sets. Moreover, the initial condition $(\mbf{x}_0,\mbf{u}_0,\mbf{w}_0,\mbf{v}_{0})$ is assumed to be feasible, i.e., consistent with the static relations in \eqref{eq:system}, and the output $\mbf{y}_0$ is computed as $\mbf{y}_0 = \mbf{C} \mbf{x}_0 + \mbf{D} \mbf{u}_{0} + \mbf{D}_v \mbf{v}_{0}$. 


Constrained zonotopes are an extension of zonotopes, proposed in \cite{Scott2016}, capable of describing also asymmetric convex polytopes, while maintaining the well-known computational benefits of zonotopes.

\begin{definition} \rm \label{def:czonotopes}
	A set $Z \subset \realset^n$ is a \emph{constrained zonotope} if $\exists (\mbf{G}_z,\mbf{c}_z,\mbf{A}_z,\mbf{b}_z) \in \realsetmat{n}{n_g} \times \realset^n \times \realsetmat{n_c}{n_g} \times \realset^{n_c}$ such that
	\normalsize
	\begin{equation} \label{eq:czdefinition}
	Z = \left\{ \mbf{c}_z + \mbf{G}_z \bm{\xi} : \| \bm{\xi} \|_\infty \leq 1, \mbf{A}_z \bm{\xi} = \mbf{b}_z \right\}.
	\end{equation}	
	\normalsize
\end{definition}

Equation \eqref{eq:czdefinition} is referred as \emph{constrained generator representation} (CG-rep). Each column of $\mbf{G}_z$ is a \emph{generator}, $\mbf{c}_z$ is the \emph{center}, and $\mbf{A}_z \bm{\xi} = \mbf{b}_z$ are the \emph{constraints}. Differently from zonotopes, the linear equality constraints in \eqref{eq:czdefinition} permit a CZ to describe general convex polytopes \citep{Scott2016}. In the following, we use the shorthand notation $Z = \{\mbf{G}_z, \mbf{c}_z,\mbf{A}_z,\mbf{b}_z \}$ for CZs, $Z = \{\mbf{G}_z, \mbf{c}_z\}$ for zonotopes. In addition, $B_\infty(\mbf{A}_z,\mbf{b}_z) = \{\bm{\xi} \in \realset^{n_g} : \ninf{\bm{\xi}} \leq 1,\,  \mbf{A}_z \bm{\xi} = \mbf{b}_z \}$ and $B_\infty^{n_g} = \{\bm{\xi} \in \realset^{n_g} : \ninf{\bm{\xi}} \leq 1\}$ \textblue{denote, respectively,} the $n_g$-dimensional constrained unitary hypercube and the unitary hypercube\footnote{The superscript $n_g$ is omitted in $B_\infty(\mbf{A}_z,\mbf{b}_z)$ for simplicity.}. A few common set operations can be computed using CZs in a trivial manner. Let $Z, W \subset \realset^{n}$, $\mbf{R} \in \realset^{m \times n}$, and $Y \subset \realset^{m}$. Define the linear mapping, Minkowski sum, and generalized intersection as
\normalsize
\begin{align}
\mbf{R}Z & \triangleq \{ \mbf{R} \mbf{z} : \mbf{z} \in Z\}, \label{eq:limage}\\
Z \oplus W & \triangleq \{ \mbf{z} + \mbf{w} : \mbf{z} \in Z,\, \mbf{w} \in W\}, \label{eq:msum}\\
Z \cap_{\mbf{R}} Y & \triangleq \{ \mbf{z} \in Z : \mbf{R} \mbf{z} \in Y\}. \label{eq:intersection}
\end{align}
\normalsize
If $Z$, $W$, $Y$ are in CG-rep, then \eqref{eq:limage}--\eqref{eq:intersection} are CZs given by
\normalsize
\begin{align}
\mbf{R}Z & \!=\! \left\{ \mbf{R} \mbf{G}_z, \mbf{R} \mbf{c}_z, \mbf{A}_z, \mbf{b}_z \right\}\!, \label{eq:czlimage}\\
Z \oplus W & \!= \!\left\{ \begin{bmatrix} \mbf{G}_z \,\; \mbf{G}_w \end{bmatrix}, \mbf{c}_z + \mbf{c}_w, \begin{bmatrix} \mbf{A}_z & \bm{0} \\ \bm{0} & \mbf{A}_w \end{bmatrix}, \begin{bmatrix} \mbf{b}_z \\ \mbf{b}_w \end{bmatrix} \right\}\!, \label{eq:czmsum}\\
Z \cap_{\mbf{R}} Y & \!=\! \left\{ \begin{bmatrix} \mbf{G}_z \; \bm{0} \end{bmatrix}, \mbf{c}_z, \begin{bmatrix} \mbf{A}_z & \bm{0} \\ \bm{0} & \mbf{A}_y \\ \mbf{R} \mbf{G}_z & -\mbf{G}_y \end{bmatrix}, \begin{bmatrix} \mbf{b}_z \\ \mbf{b}_y \\ \mbf{c}_y - \mbf{R} \mbf{c}_z \end{bmatrix} \right\}\!. \label{eq:czintersection}
\end{align}
\normalsize
%
%
With zonotopes, \eqref{eq:limage} and \eqref{eq:msum} are computed efficiently, but \eqref{eq:intersection} is conservative and difficult to compute in the general case \citep{Bravo2006}. Operations \eqref{eq:limage} and \eqref{eq:msum} are computed trivially with polytopes in vertex representation, and \eqref{eq:intersection} is computed efficiently in half-space representation, but conversions between representations are extremely expensive \citep{Hagemann2015}. On the other hand, operations \eqref{eq:limage}--\eqref{eq:intersection} \textred{are} computed trivially with \textred{CZs} by \eqref{eq:czlimage}--\eqref{eq:czintersection} resulting only in a mild increase in the complexity of the CG-rep \eqref{eq:czdefinition}. Efficient methods for complexity reduction (to enclose a CZ with another one with fewer generators and constraints) are available, see \cite{Scott2016}.

\section{State estimation of descriptor systems} \label{sec:estimation}

This section presents a new method for set-based state estimation of system \eqref{eq:system}. For the sake of clarity, the prediction-update algorithm for standard discrete-time systems \textblue{is} first recalled. Consider \eqref{eq:system} with $\mbf{E} = \mbf{I}$. Given an initial condition $X_0$ and an input $\mbf{u}_k$ with $k \geq 0$, let
\begin{equation} \label{eq:initialset}
\hat{X}_0 = \{ \mathbf{x} \in X_0 : \mbf{C} \mbf{x} + \mbf{D} \mbf{u}_{0} + \mbf{D}_v \mbf{v} = \mbf{y}_0, \, \mathbf{v} \in V \}.
\end{equation}
 Then, for $k\geq1$ the prediction-update algorithm consists in computing sets $\bar{X}_k$ and $\hat{X}_k$ satisfying \citep{Scott2016}
\begin{align}
\bar{X}_k & = \{ \mbf{A} \mbf{x} + \mbf{B} \mbf{u}_{k-1} + \mbf{B}_w \mbf{w} : \mathbf{x} \in \hat{X}_{k-1}, \, \mathbf{w} \in W \}, \label{eq:prediction0}\\
\hat{X}_k & = \{ \mathbf{x} \in \bar{X}_k : \mbf{C} \mbf{x} + \mbf{D} \mbf{u}_{k} + \mbf{D}_v \mbf{v} = \mbf{y}_k, \, \mathbf{v} \in V \}, \label{eq:update0}
\end{align}
in which \eqref{eq:prediction0} is referred to as the \emph{prediction step}, and \eqref{eq:update0} as the \emph{update step}. 
The prediction step \eqref{eq:prediction0} must be reformulated when $\mbf{E}$ is not invertible. We first consider the following assumption.
\begin{assumption} \label{ass:admissible}
	There exists a known CZ $X_\text{a} = \{\mbf{G}_\text{a}, \mbf{c}_\text{a}$, $\mbf{A}_\text{a}$, $\mbf{b}_\text{a}\}  \subset \realset^n$ such that $\mbf{x}_k \in X_\text{a}$ for all $k \geq 0$.
\end{assumption}

\begin{remark} Even though $X_\text{a}$ can be arbitrarily large,  Assumption \ref{ass:admissible} is restrictive in the sense that the descriptor system \eqref{eq:system} must be stable. Unstable systems are out of the scope of \text{this} paper and will be considered in future work.
\end{remark}

 %
%

The set estimation in \cite{Scott2016} relied on \eqref{eq:czlimage}--\eqref{eq:czintersection} to compute the steps \eqref{eq:prediction0}--\eqref{eq:update0}. In order to extend this method to the case of descriptor systems, when computing $\bar{X}_k$ it is necessary to take into account the static constraints arising from the possible singularity of $\mbf{E}$. 
The proposed method is based on singular value decomposition (SVD). Let $\mbf{E} = \mbf{U} \mbf{\Sigma} \mbf{V}^T$, where $\mbf{U}$ and $\mbf{V}$ are invertible by construction. Since $\mbf{E}$ is square, then $\mbf{\Sigma}$ is also square. Without loss of generality, let $\mbf{\Sigma}$ be arranged as $\mbf{\Sigma} = \text{blkdiag}(\tilde{\mbf{\Sigma}}, \mbf{0})$,
where $\tilde{\mbf{\Sigma}} \in \realsetmat{n_z}{n_z}$ is diagonal with all the $n_z = \text{rank}(\mbf{E})$ nonzero singular values of $\mbf{E}$. Moreover, let $\mbf{z}_k = (\tilde{\mbf{z}}_k,\check{\mbf{z}}_k) = \inv{\mbf{T}} \mbf{x}_k, \; \tilde{\mbf{z}}_k \in \realset^{n_z}, \check{\mbf{z}}_k \in \realset^{n-n_z}$,
\begin{equation}\label{eq:SVDmatrices} 
\begin{aligned}
\begin{bmatrix} \tilde{\mbf{A}} \\ \check{\mbf{A}} \end{bmatrix} & = \begin{bmatrix} \tilde{\mbf{\Sigma}}^{-1} & \mbf{0} \\ \mbf{0} & \eyenoarg \end{bmatrix} \inv{\mbf{U}} \mbf{A} \mbf{T}, \\
\begin{bmatrix} \tilde{\mbf{B}} \\ \check{\mbf{B}} \end{bmatrix} & = \begin{bmatrix} \tilde{\mbf{\Sigma}}^{-1} & \mbf{0} \\ \mbf{0} & \eyenoarg \end{bmatrix} \inv{\mbf{U}} \mbf{B}, \begin{bmatrix} \tilde{\mbf{B}}_w \\ \check{\mbf{B}}_w \end{bmatrix} = \begin{bmatrix} \tilde{\mbf{\Sigma}}^{-1} & \mbf{0} \\ \mbf{0} & \eyenoarg \end{bmatrix} \inv{\mbf{U}} \mbf{B}_w,  \\
\end{aligned}
\end{equation}
with $\mbf{T} = \inv{(\mbf{V}^T)}$, $\tilde{\mbf{A}} \in \realsetmat{n_z}{n}$, $\tilde{\mbf{B}} \in \realsetmat{n_z}{n_u}$, $\tilde{\mbf{B}}_w \in \realsetmat{n_z}{n_w}$. Then, system \eqref{eq:system} can be rewritten with decoupled dynamics given by \citep{Jonckheere1988}
\begin{subequations} \label{eq:systemSVD}
\begin{align} 
\tilde{\mbf{z}}_k & = \tilde{\mbf{A}} \mbf{z}_{k-1} + \tilde{\mbf{B}} \mbf{u}_{k-1} + \tilde{\mbf{B}}_w \mbf{w}_{k-1},  \label{eq:systemSVDdynamics} \\
\mbf{0} & = \check{\mbf{A}} \mbf{z}_{k-1} + \check{\mbf{B}} \mbf{u}_{k-1} + \check{\mbf{B}}_w \mbf{w}_{k-1}, \label{eq:systemSVDconstraints} \\
\mbf{y}_k & = \mbf{C} \mbf{T} \mbf{z}_k + \mbf{D} \mbf{u}_{k} + \mbf{D}_v \mbf{v}_{k}. \label{eq:systemSVDoutput}
\end{align}
\end{subequations}
%
Consider $W = \{\mbf{G}_w, \mbf{c}_w, \mbf{A}_w, \mbf{b}_w\}$, and $\hat{X}_{0}$ given by \eqref{eq:initialset}. From \eqref{eq:systemSVDconstraints}, the state $\mbf{z}_0$ must satisfy $\check{\mbf{A}} \mbf{z}_0 + \check{\mbf{B}} \mbf{u}_0 + \check{\mbf{B}}_w \mbf{w}_0 = \mbf{0}$. This constraint can be incorporated in the CG-rep of the initial set $\hat{Z}_0$ as follows. Let $\inv{\mbf{T}} \hat{X}_{0} \triangleq \{\mbf{G}_0,\mbf{c}_0,\mbf{A}_0,\mbf{b}_0\}$. Then, $\hat{Z}_0 \triangleq \{\hat{\mbf{G}}_0, \hat{\mbf{c}}_0,\hat{\mbf{A}}_0,\hat{\mbf{b}}_0\}$, with $\hat{\mbf{G}}_0 = [\mbf{G}_0 \,\; \mbf{0}]$, $\hat{\mbf{c}}_0 = \mbf{c}_0$,
\begin{equation*}
\begin{aligned}
\hat{\mbf{A}}_0 = \begin{bmatrix} \mbf{A}_0 & \mbf{0} \\ \check{\mbf{A}} \mbf{G}_0 & \check{\mbf{B}}_w \mbf{G}_w \end{bmatrix}, \; \hat{\mbf{b}}_0 = \begin{bmatrix}  \mbf{b}_0 \\ -\check{\mbf{A}} \mbf{c}_0 - \check{\mbf{B}}_w \mbf{c}_w - \check{\mbf{B}} \mbf{u}_{0} \end{bmatrix}.
\end{aligned}
\end{equation*}
%
Note that the extra columns in $\hat{\mbf{G}}_0$ and $\hat{\mbf{A}}_0$ come from $\mbf{w}_0 \in W$. Having defined $\hat{Z}_0$, for the purpose of state estimation the static relation \eqref{eq:systemSVDconstraints} can be shifted forward to time $k$ without loss of information. By doing so, from \eqref{eq:systemSVD}, the variables $\tilde{\mbf{z}}_k$ are fully determined by \eqref{eq:systemSVDdynamics}, while $\check{\mbf{z}}_k$ are obtained a posteriori by $\check{\mbf{A}} \mbf{z}_{k} + \check{\mbf{B}} \mbf{u}_{k} + \check{\mbf{B}}_w \mbf{w}_{k} = \mbf{0}$.

Consider the set $Z_\text{a} = \inv{\mbf{T}} X_\text{a} = \{\inv{\mbf{T}}\mbf{G}_\text{a},\inv{\mbf{T}}\mbf{c}_\text{a},\mbf{A}_\text{a},\mbf{b}_\text{a}\}$, and let $\inv{\mbf{T}} \mbf{c}_\text{a} = [ \tilde{\mbf{c}}_\text{a}^T \,\; \check{\mbf{c}}_\text{a}^T ]^T$, $\inv{\mbf{T}} \mbf{G}_\text{a} = [ \tilde{\mbf{G}}_\text{a}^T \,\; \check{\mbf{G}}_\text{a}^T ]^T$. An effective enclosure of the prediction step for the descriptor system \eqref{eq:systemSVD} can be obtained in CG-rep as follows. 

\begin{lemma} \label{lem:predictionsvd}
For $k \geq 1$, assume $\mbf{z}_{k-1} \in \hat{Z}_{k-1} = \{\hat{\mbf{G}}_{k-1}, \hat{\mbf{c}}_{k-1}$, $\hat{\mbf{A}}_{k-1}, \hat{\mbf{b}}_{k-1}\}$ and $\mbf{w}_{k-1}, \mbf{w}_k \in W = \{\mbf{G}_w, \mbf{c}_w, \mbf{A}_w, \mbf{b}_w\}$. Consider \eqref{eq:systemSVD}. If Assumption \ref{ass:admissible} holds, then $\mbf{z}_k \in \bar{Z}_k = \{\bar{\mbf{G}}_k, \bar{\mbf{c}}_k, \bar{\mbf{A}}_k, \bar{\mbf{b}}_k\}$, with
\normalsize
\begin{align*}
\bar{\mbf{G}}_k & {=}\!\! \begin{bmatrix} \tilde{\mbf{A}} \hat{\mbf{G}}_{k-1} & \!\tilde{\mbf{B}}_w \mbf{G}_w\! & \mbf{0} & \mbf{0}\\
\mbf{0} & \mbf{0} & \check{\mbf{G}}_\text{a} & \mbf{0}\end{bmatrix}\!\!,
\bar{\mbf{c}}_k {=} \!\!\begin{bmatrix} \tilde{\mbf{A}} \hat{\mbf{c}}_{k-1} {+} \tilde{\mbf{B}} \mbf{u}_{k-1} {+} \tilde{\mbf{B}}_w \mbf{c}_w \\ \check{\mbf{c}}_\text{a}\end{bmatrix}\!\!, \\  
\bar{\mbf{A}}_k & = \begin{bmatrix} \multicolumn{4}{c}{\text{blkdiag}(\hat{\mbf{A}}_{k-1}, \mbf{A}_w, \mbf{A}_\text{a}, \mbf{A}_w)} \\
\check{\mbf{A}} \begin{bmatrix} \tilde{\mbf{A}} \hat{\mbf{G}}_{k-1}  \\ \mbf{0} \end{bmatrix} & \check{\mbf{A}} \begin{bmatrix} \tilde{\mbf{B}}_w \mbf{G}_w \\ \mbf{0} \end{bmatrix} & \check{\mbf{A}} \begin{bmatrix} \mbf{0} \\ \check{\mbf{G}}_\text{a} \end{bmatrix} & \check{\mbf{B}}_w \mbf{G}_w \end{bmatrix}, \\
\bar{\mbf{b}}_k & = \begin{bmatrix} [\hat{\mbf{b}}_{k-1}^T \,\; \mbf{b}_w^T \,\; \mbf{b}_\text{a}^T \,\; \mbf{b}_w^T]^T \\ -\check{\mbf{A}} \begin{bmatrix} \tilde{\mbf{A}} \hat{\mbf{c}}_{k-1} + \tilde{\mbf{B}} \mbf{u}_{k-1} + \tilde{\mbf{B}}_w \mbf{c}_w \\ \check{\mbf{c}}_\text{a}\end{bmatrix} - \check{\mbf{B}} \mbf{u}_k - \check{\mbf{B}}_w \mbf{c}_w \end{bmatrix}.                            
\end{align*}
\normalsize
\end{lemma}
\begin{pf}
	Since by assumption $(\mbf{z}_{k-1},\mbf{w}_{k-1},\mbf{w}_k) \in \hat{Z}_{k-1} \times W \times W$, there exists $(\bm{\xi}_{k-1}, \bm{\delta}_{k-1}, \bm{\delta}_{k}) \in B_\infty(\hat{\mbf{A}}_{k-1}, \hat{\mbf{b}}_{k-1}) \times B_\infty(\mbf{A}_{w}, \mbf{b}_{w}) \times B_\infty(\mbf{A}_w, \mbf{b}_w)$ such that $\mbf{z}_{k-1} = \hat{\mbf{c}}_{k-1} + \hat{\mbf{G}}_{k-1} \bm{\xi}_{k-1}$, $\mbf{w}_{k-1} = \mbf{c}_w + \mbf{G}_w \bm{\delta}_{k-1}$, and $\mbf{w}_{k} = \mbf{c}_w + \mbf{G}_w \bm{\delta}_{k}$. Moreover, Assumption \ref{ass:admissible} implies that $\mbf{z}_k \in Z_\text{a}$. Thus there must exist $\bm{\xi}_\text{a} \in B_\infty (\mbf{A}_\text{a}, \mbf{b}_\text{a})$ such that $\check{\mbf{z}}_{k} = \check{\mbf{c}}_\text{a} + \check{\mbf{G}}_\text{a}\bm{\xi}_\text{a}$. Therefore, substituting these equalities in \eqref{eq:systemSVDdynamics} leads to
	\begin{equation} \label{eq:lema1proof1} 
	\begin{aligned}
(\tilde{\mbf{z}}_{k}, \check{\mbf{z}}_{k}) = & (\tilde{\mbf{A}} \hat{\mbf{c}}_{k-1} {+} \tilde{\mbf{B}} \mbf{u}_{k-1} {+} \tilde{\mbf{B}}_w \mbf{c}_{w} \\ & {+} \tilde{\mbf{A}} \hat{\mbf{G}}_{k-1} \bm{\xi}_{k-1} {+} \tilde{\mbf{B}}_w \mbf{G}_{w} \bm{\delta}_{k-1}, \check{\mbf{c}}_\text{a} + \check{\mbf{G}}_\text{a} \bm{\xi}_\text{a}).
    \end{aligned}
	\end{equation}
	From the constraint \eqref{eq:systemSVDconstraints} shifted to time $k$, we have that
	\begin{equation} \label{eq:lema1proof2}
	\begin{aligned}
		& \check{\mbf{B}}_w \mbf{c}_w +  \check{\mbf{B}}_w \mbf{G}_w \bm{\delta}_k + \check{\mbf{A}} \begin{bmatrix} \tilde{\mbf{A}} \hat{\mbf{c}}_{k-1} + \tilde{\mbf{B}} \mbf{u}_{k-1} + \tilde{\mbf{B}}_w \mbf{c}_{w}  \\  \check{\mbf{c}}_\text{a} \end{bmatrix} \\ & + \check{\mbf{A}} \begin{bmatrix} \tilde{\mbf{A}} \hat{\mbf{G}}_{k-1} & \tilde{\mbf{B}}_w \mbf{G}_{w} & \mbf{0} \\ \mbf{0} & \mbf{0} & \check{\mbf{G}}_\text{a} \end{bmatrix} \begin{bmatrix} \bm{\xi}_{k-1} \\ \bm{\delta}_{k-1} \\ \bm{\xi}_\text{a}\end{bmatrix} + \check{\mbf{B}} \mbf{u}_{k} = \mbf{0}.
	\end{aligned}		
	\end{equation}
	Rearranging \eqref{eq:lema1proof1} and \eqref{eq:lema1proof2}, grouping $(\bm{\xi}_{k-1}, \bm{\delta}_{k-1}, \bm{\xi}_\text{a},$ $ \bm{\delta}_{k})$, and writing in the CG-rep \eqref{eq:czdefinition} proves the lemma. \qed
\end{pf}

Lemma \ref{lem:predictionsvd} provides a predicted enclosure of the state $\mbf{z}_k$ in which the equality constraints \eqref{eq:systemSVDconstraints}, shifted to time $k$, are directly taken into account. This is possible thanks to the fact that CZs incorporate equality constraints (see \eqref{eq:czdefinition}). %
Finally, the prediction-update algorithm proposed for descriptor systems consists in the computation of CZs $\bar{Z}_k$, $\hat{Z}_k$, and $\hat{X}_k$, such that
\begin{align}
\bar{Z}_k & = \{\bar{\mbf{G}}_k, \bar{\mbf{c}}_k, \bar{\mbf{A}}_k, \bar{\mbf{b}}_k\}, \label{eq:predictionSVD} \\
\hat{Z}_k & = \bar{Z}_k \cap_{\mbf{C}\mbf{T}} ((\mbf{y}_k - \mbf{D}_u \mbf{u}_k) \oplus (-\mbf{D}_v V_k)), \label{eq:updateSVD}  \\
\hat{X}_k & = \mbf{T} \hat{Z}_k. \label{eq:finalSVD}
\end{align}
%
%
For this algorithm, \textred {the initial set is} $\hat{Z}_0$. The algorithm \eqref{eq:predictionSVD}--\eqref{eq:finalSVD} operates recursively with $\bar{Z}_k$ and $\hat{Z}_k$ for $k \geq 1$ in the transformed state-space \eqref{eq:systemSVD}, while the estimated enclosure in the original state-space \eqref{eq:system} is given by $\hat{X}_k$.

\begin{remark}
    The set $Z_\text{a}$ is used only to predict an enclosure for the components $\check{\mbf{z}}_k$. This way, the static relation \eqref{eq:systemSVDconstraints} is incorporated as constraints to the variables $\bm{\xi}_\text{a}$ in $Z_\text{a}$.
\end{remark}

\begin{remark}
	By construction, the CG-rep \eqref{eq:predictionSVD} corresponds to the exact feasible state set of \eqref{eq:systemSVD} at $k$ for the known state and uncertainty bounds. In addition, \eqref{eq:updateSVD}--\eqref{eq:finalSVD} can be computed exactly. However, in practice, in order to limit the complexity of the resulting sets these are outer approximated by using order reduction algorithms. In this case, equalities \eqref{eq:predictionSVD}-\eqref{eq:finalSVD} are replaced by the relation $\supset$. 
\end{remark}

\section{Active fault diagnosis} \label{sec:faultisolation}


The previous section presented a method to address the problem of the set-based estimation of descriptor systems. In the following, this tool is used in the design of an AFD method accounting for a finite number of possible abrupt faults. %
%
%
Consider a linear discrete-time descriptor system whose dynamics obeys one of possible $n_m$ known models
\begin{equation}
\begin{aligned} \label{eq:systemfault}
\mbf{E}^{[i]} \mbf{x}_k^{[i]} & = \mbf{A}^{[i]}  \mbf{x}_{k-1}^{[i]} + \mbf{B}^{[i]}  \mbf{u}_{k-1} + \mbf{B}_w^{[i]}  \mbf{w}_{k-1}, \\
\mbf{y}_k^{[i]} & = \mbf{C}^{[i]}  \mbf{x}_k^{[i]} + \mbf{D}^{[i]} \mbf{u}_{k} + \mbf{D}_v^{[i]} \mbf{v}_{k},
\end{aligned}
\end{equation}
for $k \geq 1$, with $\mbf{E}^{[i]} \in \realsetmat{n}{n}$, $\mbf{A}^{[i]} \in \realsetmat{n}{n}$, $\mbf{B}^{[i]} \in \realsetmat{n}{n_u}$, $\mbf{B}_w^{[i]} \in \realsetmat{n}{n_w}$, $\mbf{C}^{[i]} \in \realsetmat{n_y}{n}$, $\mbf{D}^{[i]} \in \realsetmat{n_y}{n_u}$, and $\mbf{D}_v^{[i]} \in \realsetmat{n_y}{n_v}$, $i \in \mathbb{I} = \{1,2,\ldots,n_m\}$. Moreover, $\text{rank}(\mbf{E}^{[i]}) \leq n$, and let $\mbf{x}_0^{[i]} \in X_0$, $(\mbf{w}_{k}, \mbf{v}_k) \in W \times V$, and $\mbf{u}_k \in U$, with $X_0$, $W$, $V$ and $U$ being known polytopic sets. %


The goal of AFD is to find  which model describes the process behaviour. In the following, the dynamics are assumed to not change during the diagnosis procedure, i.e. the AFD is fast enough to avoid the switching between models. In this sense, a sequence $(\mbf{u}_0, \mbf{u}_1, ..., \mbf{u}_N)$ of minimal length $N$ is designed such that any output $\mbf{y}_N^{[i]}$ is consistent with only one $i \in \mathbb{I}$. If feasible, this problem may admit multiple solutions. For this reason, we introduce a cost function and select among the feasible input sequences the optimal one. 


Let $\seq{\mbf{u}} = (\mbf{u}_0, ..., \mbf{u}_N)\in \mathbb{R}^{(N+1)n_u}$, $\seq{\mbf{w}} = (\mbf{w}_0, \ldots$, $\mbf{w}_N)\in \mathbb{R}^{(N+1)n_w}$, and $\seq{W} = W \times \ldots \times W$. Consider a variable transformation similar to the one used in the previous section. With a slight abuse of notation, let $\mbf{z}_k = (\tilde{\mbf{z}}_k,\check{\mbf{z}}_k) = (\inv{\mbf{T}} \mbf{x}_k, \mbf{w}_k),$ $\tilde{\mbf{z}}_k \in \realset^{n_z}$, $\check{\mbf{z}}_k \in \realset^{n+n_w-n_z}$,
with $\mbf{T}^{[i]} = \inv{((\mbf{V}^{[i]})^T)}$, $\mbf{V}^{[i]}$ being obtained from the SVD $\mbf{E}^{[i]} = \mbf{U}^{[i]} \mbf{\Sigma}^{[i]} (\mbf{V}^{[i]})^T$. Then, \eqref{eq:systemfault} can be rewritten as
\begin{subequations} \label{eq:systemSVDfault}
\begin{align} 
\tilde{\mbf{z}}_k^{[i]} & = \tilde{\mbf{A}}_z^{[i]} \mbf{z}_{k-1}^{[i]} + \tilde{\mbf{B}}^{[i]} \mbf{u}_{k-1}, \label{eq:systemSVDfaultdynamics} \\
\mbf{0} & = \check{\mbf{A}}_z^{[i]} \mbf{z}_{k}^{[i]} + \check{\mbf{B}}^{[i]} \mbf{u}_{k}, \label{eq:systemSVDfaultconstraints} \\
\mbf{y}_k^{[i]} & = \mbf{F}^{[i]} \mbf{z}_k^{[i]} + \mbf{D}^{[i]} \mbf{u}_{k} + \mbf{D}_v^{[i]} \mbf{v}_{k}, \label{eq:systemSVDfaultoutput}
\end{align}
\end{subequations}
with $\mbf{F}^{[i]} = \mbf{C}^{[i]} \mbf{T}^{[i]} \mbf{P}$, where $\mbf{P} = [ \eye{n} \,\; \zeros{n}{n_w}]$, $\tilde{\mbf{A}}_z^{[i]} = [\tilde{\mbf{A}}^{[i]} \,\; \tilde{\mbf{B}}_w^{[i]}]$, and $\check{\mbf{A}}_z^{[i]} = [\check{\mbf{A}}^{[i]} \,\; \check{\mbf{B}}_w^{[i]}]$. Note that the $\tilde{(\cdot)}$ and $\check{(\cdot)}$ variables are defined according to \eqref{eq:SVDmatrices} for each $i$, and equation \eqref{eq:systemSVDfaultconstraints} has been already shifted to time $k$. 

%
%
For each model $i$, consider the CZ $Z_\text{a}^{[i]} = \inv{(\mbf{T}^{[i]})} X_\text{a} \times W = \{ \mbf{G}_\text{a}^{[i]}, \mbf{c}_\text{a}^{[i]}, \mbf{A}_\text{a}^{[i]}, \mbf{b}_\text{a}^{[i]}\}$, where $X_\text{a}$ satisfies Assumption 1, the set $ \{\mbf{G}_z^{[i]},\mbf{c}_z^{[i]},\mbf{A}_z^{[i]},\mbf{b}_z^{[i]}\} = \inv{(\mbf{T}^{[i]})} X_0 \times W $, and define the initial feasible set $Z_0^{[i]} (\mbf{u}_0) = \{\mbf{z} \in \inv{(\mbf{T}^{[i]})} X_0 \times W : \eqref{eq:systemSVDfaultconstraints} \text{ holds for }k=0\}$. This set is given by $Z_0^{[i]}(\mbf{u}_0) = \{\mbf{G}_0^{[i]},\mbf{c}_0^{[i]},\mbf{A}_0^{[i]},$ $\mbf{b}_0^{[i]}(\mbf{u}_0)\}$, where $\mbf{G}_0^{[i]} = \mbf{G}_z^{[i]}$, $\mbf{c}_0^{[i]} = \mbf{c}_z^{[i]}$,
\begin{equation} \label{eq:initialAb}
\mbf{A}_0^{[i]} = \begin{bmatrix} \mbf{A}_z^{[i]} \\ \check{\mbf{A}}_z^{[i]} \mbf{G}_{0}^{[i]} \end{bmatrix}, \; \mbf{b}_0^{[i]}(\mbf{u}_0) = \begin{bmatrix} \mbf{b}_z^{[i]} \\ -\check{\mbf{A}}_z^{[i]} \mbf{c}_0^{[i]} - \check{\mbf{B}}^{[i]} \mbf{u}_{0} \end{bmatrix}.
\end{equation}
In addition, define the solution mappings $(\bm{\phi}_k^{[i]},\bm{\psi}_k^{[i]}) : \realset^{(k+1)n_u} \times \realset^n \times \realset^{(k+1)n_w} \times \realset^{n_v} \to \realset^{n+n_w} \times \realset^{n_y}$ such that $\bm{\phi}_k^{[i]}(\seq{\mbf{u}},\mbf{z}_0, \seq{\mbf{w}})$ and $\bm{\psi}_k^{[i]}(\seq{\mbf{u}},\mbf{z}_0, \seq{\mbf{w}}, \mbf{v}_k)$ are the state and output of \eqref{eq:systemSVDfault} at $k$, respectively. Then, for each $i \in \mathbb{I}$, define \emph{state and output reachable sets} at time $k$ as 
\begin{equation*}
\begin{aligned}
Z_k^{[i]}(\seq{\mbf{u}}) {=} & \{ \bm{\phi}_k^{[i]}(\seq{\mbf{u}},\mbf{z}_0, \seq{\mbf{w}}) : (\mbf{z}_0^{[i]}, \seq{\mbf{w}}) \in Z_0^{[i]}(\mbf{u}_0) \times \seq{W} \},\\
Y_k^{[i]}(\seq{\mbf{u}}) {=} & \{ \bm{\psi}_k^{[i]}(\seq{\mbf{u}},\mbf{z}_0, \seq{\mbf{w}}, \mbf{v}_k) \!:\! (\mbf{z}_0,\seq{\mbf{w}},\mbf{v}_k) \!\in\! Z_0^{[i]}(\mbf{u}_0){\times} \seq{W} {\times} V\}.
\end{aligned}
\end{equation*}
%
Using \eqref{eq:czlimage}--\eqref{eq:czmsum}, and taking note that by assumption $\mbf{z}_k^{[i]} \in Z_\text{a}^{[i]} \subset \realset^n \times W$ for every $k \geq 0$, the set $Z_N^{[i]}(\seq{\mbf{u}})$ is given by the CZ $\{ \mbf{G}_N^{[i]}, \mbf{c}_N^{[i]}(\seq{\mbf{u}}), \mbf{A}_N^{[i]}, \mbf{b}_N^{[i]}(\seq{\mbf{u}})\},$ 
where $\mbf{G}_N^{[i]}$, $\mbf{c}_N^{[i]}(\seq{\mbf{u}})$, $\mbf{A}_N^{[i]}$, and $\mbf{b}_N^{[i]}(\seq{\mbf{u}})$ are obtained by the recursive relations
\normalsize
\begin{align*}
& \mbf{c}_k^{[i]}(\seq{\mbf{u}}) {=} \!\begin{bmatrix} \tilde{\mbf{A}}_z^{[i]} \mbf{c}_{k-1}^{[i]}(\seq{\mbf{u}}) + \tilde{\mbf{B}}^{[i]} \mbf{u}_{k-1} \\ \check{\mbf{c}}_\text{a}^{[i]} \end{bmatrix}\!, \mbf{G}_k^{[i]} {=} \! \begin{bmatrix} \tilde{\mbf{A}}_z^{[i]} \mbf{G}_{k-1}^{[i]}\!\! & \mbf{0} \\ \mbf{0} & \check{\mbf{G}}_\text{a}^{[i]}\end{bmatrix}\!, \\
& \mbf{A}_k^{[i]} = \begin{bmatrix} \mbf{A}_{k-1}^{[i]} & \mbf{0} \\ \mbf{0} & \mbf{A}_\text{a}^{[i]} \\ \multicolumn{2}{c}{\check{\mbf{A}}_z^{[i]} \mbf{G}_{k}^{[i]}} \end{bmatrix}, \mbf{b}_k^{[i]}(\seq{\mbf{u}}) = \begin{bmatrix} \mbf{b}_{k-1}^{[i]}(\seq{\mbf{u}}) \\  \mbf{b}_\text{a}^{[i]} \\ -\check{\mbf{A}}_z^{[i]} \mbf{c}_k^{[i]}(\seq{\mbf{u}}) - \check{\mbf{B}}^{[i]} \mbf{u}_{k} \end{bmatrix},  
\end{align*}\color{black}
\normalsize
for $k = 1,2,\ldots,N$. %
Note that the third constraint in $\mbf{A}_k^{[i]}$, $\mbf{b}_k^{[i]}(\seq{\mbf{u}})$, comes from the fact that \eqref{eq:systemSVDfaultconstraints} must hold. 
%
%
Using the initial values \eqref{eq:initialAb}, \textred{the variables $\mbf{c}_N^{[i]}(\seq{\mbf{u}})$ and $\mbf{b}_N^{[i]}(\seq{\mbf{u}})$} can be written as explicit functions of the input sequence $\seq{\mbf{u}}$ as
\begin{align}
	\mbf{c}_N^{[i]}(\seq{\mbf{u}}) & {=} \begin{bmatrix} \tilde{\mbf{A}}_z^{[i]} \\ \mbf{0} \end{bmatrix}^{N}\!\!\!\! \mbf{c}_z^{[i]} {+} \!\!\sum_{m=1}^{N} \!\! \bigg( \begin{bmatrix} \tilde{\mbf{A}}_z^{[i]} \\ \mbf{0} \end{bmatrix}^{m\!-\!1} \!\! \begin{bmatrix} \mbf{0} \\ \check{\mbf{c}}_\text{a}^{[i]} \end{bmatrix} \bigg) {+} \mbf{H}_N^{[i]} \seq{\mbf{u}} \label{eq:stateck},\\
\mbf{b}_N^{[i]}(\seq{\mbf{u}}) & = \bm{\alpha}_N^{[i]} + \bm{\Lambda}_N^{[i]} \mbf{c}_z^{[i]} + \bm{\Omega}_N^{[i]} \seq{\mbf{u}}, \label{eq:statebk} 
\end{align}
where $\bm{\alpha}_N^{[i]} = \bm{\beta}_N^{[i]} + \bm{\Upsilon}_N^{[i]} \mbf{p}_N^{[i]}$, $\bm{\Lambda}_N^{[i]} = \bm{\Upsilon}_N^{[i]} \mbf{Q}_N^{[i]}$, $\bm{\Omega}_N^{[i]} = \bm{\Gamma}_N^{[i]}+ \bm{\Upsilon}_N^{[i]} \seq{\mbf{H}}\zerospace^{[i]}$, $\textred{\bm{\beta}_N^{[i]} = \begin{bmatrix} [(\mbf{b}_z^{[i]})^T \; \mbf{0}] & [(\mbf{b}_\text{a}^{[i]})^T \; \mbf{0}] & \cdots & [ (\mbf{b}_\text{a}^{[i]})^T \; \mbf{0} ] \end{bmatrix}^T,}$
\begin{align*}
& \mbf{p}_N^{[i]} = \bigg[ \begin{matrix} \begin{bmatrix} \mbf{0} \\ \mbf{0} \end{bmatrix}^T & \begin{bmatrix} \mbf{0} \\ \check{\mbf{c}}_\text{a}^{[i]} \end{bmatrix}^T & \cdots & \sum_{m=1}^{N} \bigg( \begin{bmatrix} \tilde{\mbf{A}}_z^{[i]} \\ \mbf{0} \end{bmatrix}^{m-1} \begin{bmatrix} \mbf{0} \\ \check{\mbf{c}}_\text{a}^{[i]} \end{bmatrix} \bigg)^T \end{matrix} \bigg]^T, \\
& \textred{\bm{\Upsilon}_N^{[i]} = \text{blkdiag} \big( \big[\mbf{0} \,\; (-\check{\mbf{A}}_z^{[i]})^T\big]^T , \ldots, \big[\mbf{0} \,\; (-\check{\mbf{A}}_z^{[i]})^T\big]^T \big),} \\
& \textred{\bm{\Gamma}_N^{[i]} = \text{blkdiag} \big(\big[\mbf{0} \,\; (-\check{\mbf{B}}^{[i]})^T\big]^T, \ldots, \big[\mbf{0} \,\; (-\check{\mbf{B}}^{[i]})^T\big]^T \big),} \\
& \mbf{Q}_N^{[i]} = \bigg[\begin{matrix} \cdots & \bigg(\begin{bmatrix} \tilde{\mbf{A}}_z^{[i]} \\ \mbf{0} \end{bmatrix}^\ell\bigg)^T & \cdots \end{matrix}\bigg]^T\!\!, \; \seq{\mbf{H}}\zerospace^{[i]} = \big[\begin{matrix} \cdots & (\mbf{H}_\ell^{[i]})^T & \cdots \end{matrix}\big]^T\!\!, \\
& \mbf{H}_h^{[i]} = \left[\begin{matrix} \cdots & \underbrace{\begin{bmatrix} \tilde{\mbf{A}}_z^{[i]} \\ \mbf{0} \end{bmatrix}^{h-m} \begin{bmatrix} \tilde{\mbf{B}}^{[i]} \\ \mbf{0} \end{bmatrix}}_{m = 1,2,\ldots,h} & \cdots & \underbrace{\begin{bmatrix} \mbf{0} \\ \mbf{0} \end{bmatrix}}_{N-h+1 \text{ terms}} & \cdots \end{matrix} \right], 
\end{align*}
with $\ell = 0,\ldots,N$. The variables $\bm{\beta}_N$, $\mbf{p}_N$, $\bm{\Upsilon}_N$, and $\bm{\Gamma}_N$ have $N+1$ block matrices. In addition, the expression $\mbf{H}_h^{[i]}$ holds for $h = 1,\ldots,N$, while $\mbf{H}_0^{[i]} = \zeros{n}{(N{+}1)n_u}$. 

Since $Z_N^{[i]}(\seq{\mbf{u}})$ is a \textred{CZ}, the output reachable set $Y_N^{[i]}(\seq{\mbf{u}})$ is then a \textred{CZ} obtained in accordance with \eqref{eq:systemSVDfaultoutput} as $Y_N^{[i]}(\seq{\mbf{u}}) = \mbf{F}^{[i]}Z_N^{[i]} (\seq{\mbf{u}}) \oplus \mbf{D}^{[i]}\mbf{u}_N \oplus \mbf{D}_v^{[i]} V.$ Using properties \eqref{eq:czlimage} and \eqref{eq:czmsum}, and letting $V = \{ \mbf{G}_v, \mbf{c}_v, \mbf{A}_v,$ $ \mbf{b}_v\}$, this set is $Y_N^{[i]}(\seq{\mbf{u}}) = \{ \mbf{G}_N^{Y[i]}, \mbf{c}_N^{Y[i]}(\seq{\mbf{u}})$, $\mbf{A}_N^{Y[i]},$ $ \mbf{b}_N^{Y[i]}(\seq{\mbf{u}})\}$, with
\begin{subequations} \label{eq:outputreachable}
\begin{align}
& \mbf{c}_N^{Y[i]}(\seq{\mbf{u}}) = \mbf{F}^{[i]} \mbf{c}_N^{[i]}(\seq{\mbf{u}}) + \mbf{D}^{[i]}\mbf{u}_N + \mbf{D}_v^{[i]} \mbf{c}_v, \label{eq:outputck}\\
& \mbf{G}_N^{Y[i]} = [ \mbf{F}^{[i]} \mbf{G}_N^{[i]} \,\; \mbf{G}_v ],\; \mbf{A}_N^{Y[i]} = \text{blkdiag}(\mbf{A}_N^{[i]},\mbf{A}_v), \label{eq:outputGkAk}\\
& \mbf{b}_N^{Y[i]}(\seq{\mbf{u}}) = [(\mbf{b}_N^{[i]}(\seq{\mbf{u}}))^T \,\; \mbf{b}_v^T ]^T \label{eq:outputbk}. 
\end{align}
\end{subequations}

%
Consider an input sequence $\seq{\mbf{u}} \in \seq{U}$ to be injected into the \textgreen{set of models} \eqref{eq:systemSVDfault}, and let $\mbf{y}^{[i]}_N$ denote the observed output. We are interested in the design of an input sequence such that the relation $\mbf{y}_N^{[i]} \in Y_N^{[i]}(\seq{\mbf{u}})$ 
is valid for only one $i \in \mathbb{I}$.
\begin{definition} \rm \label{def:separatinginput}
	\textred{An input sequence $\seq{\mbf{u}}$ is said to be a \emph{separating input} on $k \in [0, N]$ if, for every $i,j \in \mathbb{I}$, $i \neq j$},
	\begin{equation} \label{eq:separatinginputdef}
	Y_N^{[i]}(\seq{\mbf{u}}) \cap Y_N^{[j]}(\seq{\mbf{u}}) = \emptyset.
	\end{equation}
\end{definition}
Clearly, if \eqref{eq:separatinginputdef} holds for \textred{all} $i,j \in \mathbb{I}$, $i \neq j$, then $\mbf{y}_N^{[i]} \in Y_N^{[i]}(\seq{\mbf{u}})$ must hold only for one $i$. In the case that this is not valid for any $i \in \mathbb{I}$, one concludes that the real dynamics does not belong to the set of models \eqref{eq:systemSVDfault}. \textred{The following theorem is based on the computation of $Y_N^{[i]}(\seq{\mbf{u}})$ expressed by \eqref{eq:outputreachable} and the results in \cite{Raimondo2016}.}

\begin{theorem} \rm \label{thm:separatinginput}
	\textred{An input} $\seq{\mbf{u}} \in \seq{U}$ is a separating input iff
	\begin{equation} \label{eq:separationconditioncz}
	\begin{bmatrix} \mbf{N}(i,j) \\ \bm{\Omega}(i,j) \end{bmatrix} \seq{\mbf{u}} \!\notin\! \mathcal{Y}(i,j) \!=\! \left\{ \begin{bmatrix} \mbf{G}^Y_N(i,j) \\ \mbf{A}^Y_N(i,j) \end{bmatrix} , \begin{bmatrix} \mbf{c}^Y_N(i,j) \\ - \mbf{b}^Y_N(i,j) \end{bmatrix} \right\},
	\end{equation}
    $\forall i,j \in \mathbb{I}$, $i \neq j$, where $\mbf{N}(i,j) = \mbf{F}^{[j]} \seq{\mbf{H}}\zerospace^{[j]} - \mbf{F}^{[i]} \seq{\mbf{H}}\zerospace^{[i]} + [\mbf{0} \,\; (\mbf{D}^{[j]}-\mbf{D}^{[i]})]$, $\bm{\Omega}(i,j) = [(\bm{\Omega}_N^{[i]})^T \; \mbf{0} \; (\bm{\Omega}_N^{[j]})^T \; \mbf{0}]^T$, and 
	\begin{align*}
	& \mbf{G}^Y_N(i,j) = [ \mbf{G}^{Y[i]}_N \; {-}\mbf{G}^{Y[j]}_N ],\, \mbf{c}^Y_N(i,j) = \mbf{c}^{Y[i]}_N(\seq{\mbf{0}}) - \mbf{c}^{Y[j]}_N(\seq{\mbf{0}}),\\
	& \mbf{A}^Y_N(i,j) = \begin{bmatrix} \mbf{A}_N^{Y[i]} & \mbf{0} \\ \mbf{0}  & \mbf{A}_N^{Y[j]}  \end{bmatrix},\; \mbf{b}^Y_N(i,j) = \begin{bmatrix} \mbf{b}_N^{Y[i]}(\seq{\mbf{0}}) \\ \mbf{b}_N^{Y[j]}(\seq{\mbf{0}}) \end{bmatrix},
	\end{align*}
	where $\seq{\mbf{0}}$ denotes the zero input sequence.
\end{theorem}

\begin{pf}
	\textred{The relations below follow from \eqref{eq:stateck}--\eqref{eq:outputck}, \eqref{eq:outputbk}:}
	\begin{align}
	\mbf{c}^{Y[i]}_N(\seq{\mbf{u}}) & = \mbf{c}^{Y[i]}_N(\seq{\mbf{0}}) + \mbf{F}^{[i]} \mbf{H}_N^{[i]} \seq{\mbf{u}} + \mbf{D}^{[i]} \mbf{u}_N, \label{eq:outputck0}\\
	\mbf{b}^{Y[i]}_N(\seq{\mbf{u}}) & = \mbf{b}^{Y[i]}_N(\seq{\mbf{0}}) + [ (\bm{\Omega}_N^{[i]})^T \; \mbf{0}]^T \seq{\mbf{u}}.\label{eq:outputbk0}
	\end{align}
	\textred{From \eqref{eq:czintersection}} with $\mbf{R} = \mbf{I}$, \textred{\eqref{eq:separatinginputdef}} is true iff $\nexists \bm{\xi} \in B_\infty$ such that
	\begin{equation*}
	\begin{bmatrix} \mbf{A}_N^{Y[i]} & \mbf{0} \\ \mbf{0} & \mbf{A}_N^{Y[j]} \\ \mbf{G}_N^{Y[i]} & - \mbf{G}_N^{Y[j]} \end{bmatrix} \bm{\xi} = \begin{bmatrix} \mbf{b}_N^{Y[i]}(\seq{\mbf{u}}) \\ \mbf{b}_N^{Y[j]}(\seq{\mbf{u}}) \\ \mbf{c}_N^{Y[j]}(\seq{\mbf{u}}) - \mbf{c}_N^{Y[i]}(\seq{\mbf{u}}) \end{bmatrix}.
	\end{equation*}
	According to \eqref{eq:outputck0}--\eqref{eq:outputbk0} one has \textred{$\mbf{c}_N^{Y[j]}(\seq{\mbf{u}}) - \mbf{c}_N^{Y[i]}(\seq{\mbf{u}}) = - \mbf{c}^Y_N(i,j) + \mbf{N}(i,j) \seq{\mbf{u}}$, $[(\mbf{b}_N^{Y[i]}(\seq{\mbf{u}}))^T  \;  (\mbf{b}_N^{Y[j]}(\seq{\mbf{u}}))^T]^T $ $ \!=\! \mbf{b}^Y_N(i,j)$ $+ \bm{\Omega}(i,j) \seq{\mbf{u}}$,} 
	with $\mbf{c}^Y_N(i,j)$, $\mbf{b}^Y_N(i,j)$, $\mbf{N}(i,j)$, and $\bm{\Omega}(i,j)$ defined as in the statement of the theorem. Then, \eqref{eq:separatinginputdef} holds iff $\nexists \bm{\xi} \in B_\infty$ such that $\mbf{G}^Y_N(i,j) \bm{\xi} = - \mbf{c}^Y_N(i,j) + \mbf{N}(i,j) \seq{\mbf{u}}$, and $\mbf{A}^Y_N(i,j) \bm{\xi} = \mbf{b}^Y_N(i,j) +  \bm{\Omega}(i,j) \seq{\mbf{u}}$. This is equivalent to
	\begin{align*}
    \nexists \bm{\xi} \in B_\infty: \begin{bmatrix} \mbf{G}^Y_N(i,j) \\ \mbf{A}^Y_N(i,j) \end{bmatrix} \bm{\xi} + \begin{bmatrix} \mbf{c}^Y_N(i,j) \\ - \mbf{b}^Y_N(i,j) \end{bmatrix} = \begin{bmatrix} \mbf{N}(i,j) \\ \bm{\Omega}(i,j) \end{bmatrix} \seq{\mbf{u}}, 
	\end{align*}
	which in turn holds iff \eqref{eq:separationconditioncz} is satisfied, with $\mbf{G}^Y_N(i,j)$ and $\mbf{A}^Y_N(i,j)$ defined as in the statement of the theorem. \qed
	%
\end{pf}

\textred{Let $n_q$ denote} the number of all possible combinations \textred{of} $i,j \in \mathbb{I}$, $i \neq j$, and define $\mbf{N}^{\mathcal{Y}[q]} = [\mbf{N}^T(i,j) \; \bm{\Omega}^T(i,j)]^T$, $\mathcal{Y}^{[q]} = \mathcal{Y}(i,j) = \{\mbf{G}^{\mathcal{Y}[q]}, \mbf{c}^{\mathcal{Y}[q]}\}$, for each $q \in \{ 1,2,\ldots,n_q\}$, with \textred{$\{\mbf{G}^{\mathcal{Y}[q]}$, $\mbf{c}^{\mathcal{Y}[q]}\}$ being the} right hand side of \eqref{eq:separationconditioncz}. As it can be noticed, $\mathcal{Y}^{[q]}$ is a zonotope. Then, \textred{the relation $\mbf{N}^{\mathcal{Y}[q]} \seq{\mbf{u}} \notin \mathcal{Y}^{[q]}$ can be verified by solving} a linear program (LP) similar to what proposed in \cite{Scott2014}. In this sense, \textred{the following lemma provides an effective way to verify} if a given input sequence is a separating input according to Theorem \ref{thm:separatinginput}, \textred{consequently satisfying \eqref{eq:separatinginputdef}}.

\begin{lemma} \rm \label{lem:verifyseparating}
	Let $\mathcal{Y}^{[q]} = \{\mbf{G}^{\mathcal{Y}[q]}, \mbf{c}^{\mathcal{Y}[q]}\}$. For each $\seq{\mbf{u}} \in \seq{U}$ and $q \in \{1,2,\ldots,n_q\}$, define \textred{$\hat{\delta}^{[q]}(\seq{\mbf{u}}) = \underset{\delta^{[q]}, \bm{\xi}^{[q]}}{\min} \delta^{[q]}$, subject to}
	\begin{equation*}
\mbf{N}^{\mathcal{Y}[q]} \seq{\mbf{u}} = \mbf{G}^{\mathcal{Y}[q]} \bm{\xi}^{[q]} + \mbf{c}^{\mathcal{Y}[q]}, \quad \ninf{\bm{\xi}} \leq 1 + \delta^{[q]}.
	\end{equation*}
	\color{black}
	Then $\mbf{N}^{\mathcal{Y}[q]} \seq{\mbf{u}} \notin \mathcal{Y}^{[q]} \iff \hat{\delta}^{[q]}(\seq{\mbf{u}}) > 0$.
\end{lemma}
\begin{pf}
	See Lemma 4 in \cite{Scott2014}.
\end{pf}

For the \textred{AFD} of \textgreen{the $n_m$ models in \eqref{eq:systemSVDfault} of the descriptor system}, we consider the design of \textred{a} separating input of minimum length according to the optimization problem
\begin{equation} \label{eq:optimalseparatingcz}
\underset{\seq{\mbf{u}} \in \seq{U}}{\min} ~ \{J(\seq{\mbf{u}})
: \mbf{N}^{\mathcal{Y}[q]} \seq{\mbf{u}} \notin \mathcal{Y}^{[q]}, ~ \forall q  = 1,2,\ldots,n_q\},   
\end{equation}
\textred{with $J(\seq{\mbf{u}})$ chosen} to minimize any harmful effects caused by \textred{injecting $\seq{\mbf{u}}$ into \eqref{eq:systemSVDfault}}. For simplicity we may choose $J(\seq{\mbf{u}}) = \sum_{j=0}^{N} \mbf{u}_{j}^T \mbf{R} \mbf{u}_{j}$, \textred{where $\mbf{R}$ is a weighting matrix}. \textred{As in \cite{Scott2014}, this is a bilevel optimization problem and can be rewritten as a mixed-integer quadratic program} by defining a \emph{minimum separation threshold} $\varepsilon > 0$ such that $\varepsilon \leq \hat{\delta}^{[q]}(\seq{\mbf{u}})$, for all $q  = 1,2,\ldots,n_q$.

\section{Numerical examples}\label{sec:results}


\textred{This section first evaluates} the accuracy of the \textred{state estimation method proposed in Section \ref{sec:estimation} for descriptor systems using \textred{CZs}}. %
%
Consider \textred{system} \eqref{eq:system} with matrices \textred{$\mbf{E} = \text{diag}(1,1,0)$, $\mbf{B}_w = \text{diag}(0.1,1.5,0.6)$, $\mbf{D}_v = \text{diag}(0.5,1.5)$},
\begin{equation*}
\mbf{A} = \begin{bmatrix} 0.5 & 0 & 0 \\ 0.8 & 0.95 & 0 \\ -1 & 0.5 & 1 \end{bmatrix}, \; \mbf{B} = \begin{bmatrix} 1 & 0 \\ 0 & 1 \\ 0 & 0 \end{bmatrix}, \; \mbf{C} = \begin{bmatrix} 1 & 0 & 1 \\ 1 & -1 & 0 \end{bmatrix}, 
\end{equation*}
and $\mbf{D} = \mbf{0}$. The initial state $\mbf{x}_0$ is bounded by the zonotope
\begin{equation} \label{eq:estimationx0}
X_0 = \left\{ \text{diag}(0.1, 1.5, 0.6),\, [ 0.5 \; 0.5 \; 0.25]^T \right\},
\end{equation}
and the uncertainties are random uniform noises bounded by $\|\mbf{w}_k\|_\infty \leq 1$, $\|\mbf{v}_k\|_\infty \leq 1$. The \textred{CZ} in Assumption 1 is $X_\text{a} = \{50{\cdot}\eye{3},\mbf{0}\}$. The simulation is conducted for $\mbf{x}_0 = [ 0.5 \; 0.5 \; 0.25 ]^T$, and the complexity of the \textred{CZs} is limited to 15 generators and 5 constraints using the constraint elimination algorithm in \cite{Scott2016} and Method 4 in \cite{Scott2018}.%
\begin{figure}[!tb]
	\centering{
		\def\svgwidth{0.8\columnwidth}
		{\scriptsize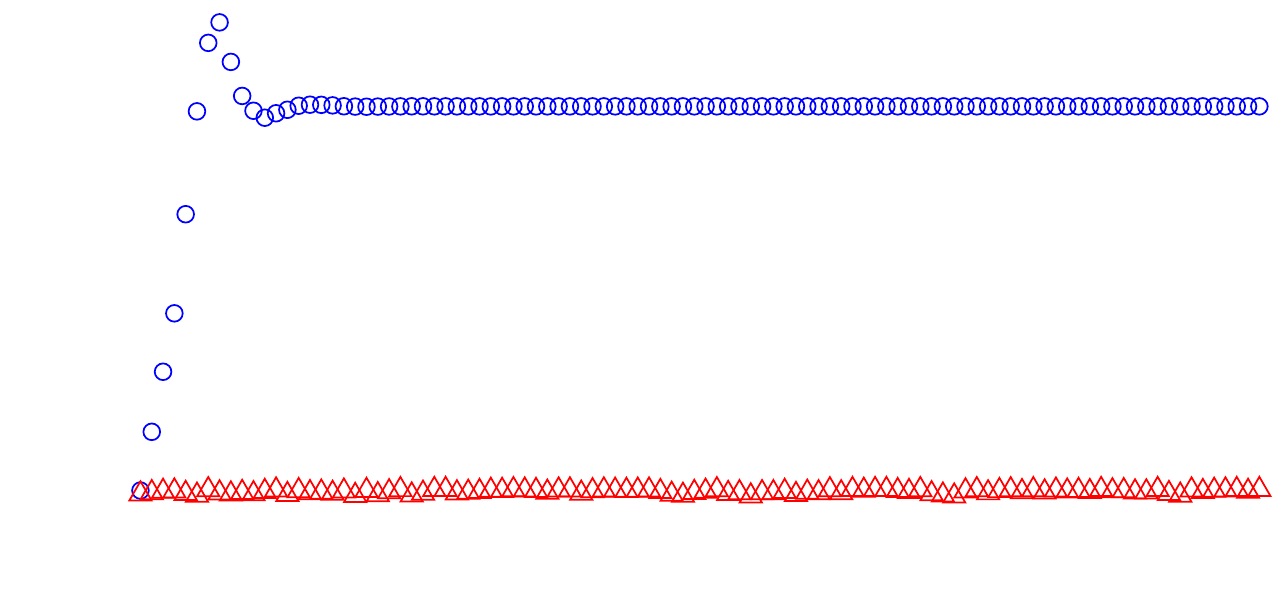}
		\def\svgwidth{0.8\columnwidth}
		{\scriptsize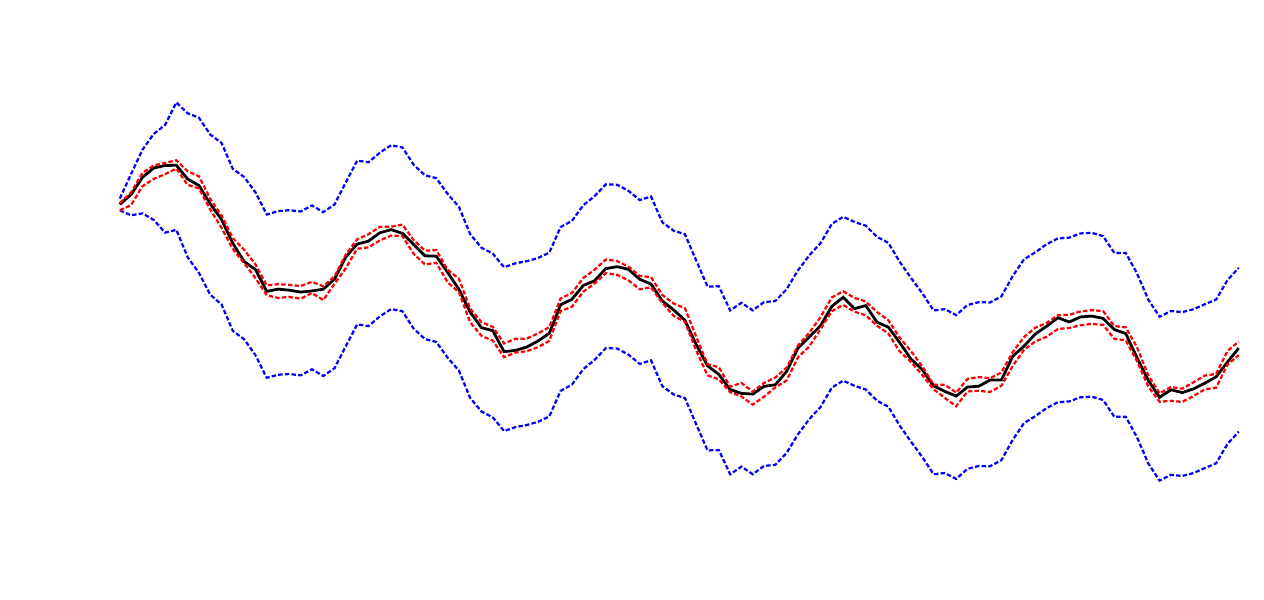}
		\vspace{-2mm}
		\caption{\textred{The volumes of $\hat{X}_k$ obtained using the estimation method in Section \ref{sec:estimation} and the zonotope method (top), as well as the projections of $\hat{X}_k$ onto $x_3$ (bottom).}}\label{fig:estimationx3}}
\end{figure}

Fig. \ref{fig:estimationx3} shows the volumes of the CZs $\hat{X}_k$ for $k \in [0,100]$ obtained using the algorithm \eqref{eq:predictionSVD}--\eqref{eq:finalSVD}. Results obtained using the zonotope method in \cite{Puig2018} are presented for comparison\footnote{Specifically, this is the set-membership approach proposed in \cite{Puig2018} with Kalman correction matrix.}. Note that the computation of the volumes was possible because this example has few states\textred{, otherwise} the radii (half the length of the longest edge of the interval hull) can be used instead. The complexity of the zonotopes is limited to 15 generators using Method 4 in \cite{Scott2018}. \textred{Fig. \ref{fig:estimationx3} shows also} the projections of $\hat{X}_k$ onto $x_3$. As it can be noticed, CZs provide substantially sharper \textred{bounds} in comparison to zonotopes. This is possible since the enclosure in Lemma \ref{lem:predictionsvd} takes into account the static constraints explicitly, while zonotopes provide only a conservative bound of the \textred{corresponding feasible region.} \textred{However}, this improved accuracy comes with an increase in computational time due to the higher set complexity (see \cite{Scott2016} for a discussion). \textred{This experiment was run 500 times consecutively on a laptop with an Intel Core i7-9750H processor, resulting in an} average execution time of 3.34 ms for \textred{CZs}, and of 0.33 ms for zonotopes.

\begin{figure*}[!htb]
	\centering{
		\def\svgwidth{0.8\textwidth}
		{\scriptsize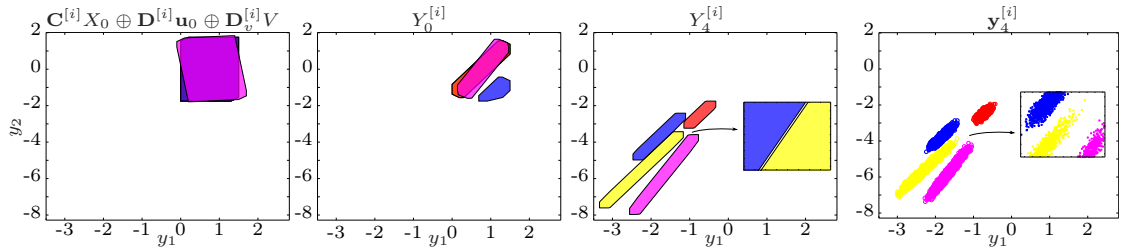}
		\vspace{-2mm}
		\caption{The sets $\mbf{C}^{[i]} X_0 \oplus \mbf{D}^{[i]} \mbf{u}_0 \oplus \mbf{D}_v^{[i]} V$, the output reachable sets $Y_0^{[i]}$, $Y_4^{[i]}$, and 2000 samples for $\mbf{y}^{[i]}_4$, with $i=1$ (yellow), $i=2$ (red), $i=3$ (blue), and $i=4$ (magenta), obtained by the injection of the input sequence $\seq{\mbf{u}}$.}\label{fig:fdi}}
\end{figure*}

We now evaluate the effectiveness of the \textred{AFD} method proposed in Section \ref{sec:faultisolation}. Consider the \textgreen{set of models} \eqref{eq:systemfault} with model $i = 1$ described in the previous example, \textred{and}
\begin{equation*}
	\begin{aligned}
	& \mbf{A}^{[2]} = \begin{bmatrix} 0.5 & 0 & 0 \\ 0.8 & 0.6 & 0 \\ -1 & 0.5 & 1 \end{bmatrix}, \; \mbf{B}^{[3]} = \begin{bmatrix} 1 & 0 \\ 0 & 0 \\ -1 & 0 \end{bmatrix}, \; \mbf{C}^{[4]} = \begin{bmatrix} 1 & 0.1 & 1 \\ 1 & -1 & 0.1 \end{bmatrix},
	\end{aligned}
\end{equation*}
$\mbf{E}^{[i]} = \mbf{E}^{[1]}$, $\mbf{B}^{[2]} = \mbf{B}^{[4]} = \mbf{B}^{[1]}$, $\mbf{C}^{[2]} = \mbf{C}^{[3]} = \mbf{C}^{[1]}$, $\mbf{B}_w^{[i]} = \mbf{B}_w^{[1]}$, and $\mbf{D}_v^{[i]} = \mbf{D}_v^{[1]}$, $\mbf{D}^{[i]} = \mbf{D}^{[1]}$, $i \in \{2,3,4\}$. The initial state $\mbf{x}_0^{[i]}$ is bounded by \eqref{eq:estimationx0}, 
the uncertainties are bounded by $\|\mbf{w}_k\|_\infty \leq 0.1$, $\|\mbf{v}_k\|_\infty \leq 0.1$, and the input is limited by $\|\mbf{u}_k\|_\infty \leq 1$. Let $X_\text{a} = \{50{\cdot}\eye{3},\mbf{0}\}$ and $\varepsilon = 0.01$. All the models $i \in \mathbb{I}$ are considered to be faulty and must be separated. The number of generators of $\mathcal{Y}(i,j)$ was limited to two times its dimension using Method 4 in \cite{Scott2018}. The minimum length optimal input sequence that solves \eqref{eq:optimalseparatingcz} was obtained using CPLEX 12.8 and MATLAB 9.1, with $J(\seq{\mbf{u}}) = \seq{\mbf{u}}\zerospace^T \seq{\mbf{u}}$, \textred{and is} 
\begin{equation*}
\seq{\mbf{u}} = \left(\begin{bmatrix} 1 \\ 1 \end{bmatrix},\begin{bmatrix} 0.73 \\ 1 \end{bmatrix}, \begin{bmatrix} 0 \\ 0.92 \end{bmatrix}, \begin{bmatrix} 0 \\ 0 \end{bmatrix}, \begin{bmatrix} -0.45 \\ 0 \end{bmatrix} \right).
\end{equation*}
%
Fig. \ref{fig:fdi} shows the output reachable sets $Y_0^{[i]}$ and $Y_4^{[i]}$ for \textred{$i=1,2,3,4$}, resulting by the injection of the designed \textred{$\seq{\mbf{u}}$}. We also show the sets defined by $\mbf{C}^{[i]} X_0 \oplus \mbf{D}^{[i]} \mbf{u}_0 \oplus \mbf{D}_v^{[i]} V$, which do not take into account the equality constraint \eqref{eq:systemSVDfaultconstraints}. Note that these sets are completely overlapped for \textred{$i \in \{1,2,3,4\}$}, and $Y_0^{[i]}$ are completely overlapped for \textred{$i \in \{1,2,4\}$}. On the other hand, $Y_4^{[i]}$ are disjoint for every $i \in \mathbb{I}$, showing that the \textred{injection of $\seq{\mbf{u}}$} guarantees fault diagnosis at $k = 4$. Fig. \ref{fig:fdi} also shows clouds containing 2000 samples of the output $\mbf{y}^{[i]}_4$ for each model $i$, demonstrating that the outputs of the \textgreen{models} \eqref{eq:systemfault} are in fact separated.
%

\section{Conclusion} \label{sec:conclusion}

This paper proposes novel algorithms for set-valued state estimation and \textred{AFD} of linear descriptor systems with \textred{unknown-but-bounded uncertainties}. \textred{The methods} \textred{use} \textred{CZs}, a generalization of zonotopes capable of describing strongly asymmetric convex sets. \textred{This leads to} significantly tighter results \textred{than zonotope approaches}. In addition, \textred{AFD} was enabled without \textred{assuming rank properties} on the structure of the \textred{system}. The effectiveness of the \textred{new} methods was \textblue{corroborated by} numerical examples.



              
                                                                         
\bibliography{thesis_bibliography,UAVControl_bib,Surveys_bib,PassiveFTC_bib,ActiveFTC_bib,UAVFTC,SetTheoretic_bib,SetTheoreticFTCFDI_bib,BackgroundHist_bib,Davide_bib,paperAutomatica_bib,paperCDC_bib,paperECC_bib,paper_bibliography,qualify_bib,Robotic_bib}

\end{document}